\documentclass[onecolumn]{IEEEtran}

\usepackage[pdftex]{graphicx}
\usepackage[psamsfonts]{amssymb}
\usepackage[utf8x]{inputenc}
\usepackage{graphicx}

\usepackage{amsmath}
\usepackage{dsfont}
\usepackage{comment}

\usepackage{cases}

\providecommand{\abs}[1]{\left\lvert#1\right\rvert}
\providecommand{\norm}[1]{\left\lVert#1\right\rVert}
\providecommand{\vc}[1]{\boldsymbol{#1}}

\title{Communication/Computation Tradeoffs in Consensus-Based Distributed Optimization}

\author{Konstantinos I. Tsianos, Sean Lawlor and Michael G. Rabbat
\thanks{K. I. Tsianos is a PhD candidate at the Department of Electrical and Computer Engineering,         McGill University, Montreal, Quebec H3A 2A7, Canada, 
        {\tt\small konstantinos.tsianos@mail.mcgill.ca}}%
\thanks{S. Lawlor is a Master's student at the Department of Electrical and Computer Engineering,         McGill University, Montreal, Quebec H3A 2A7, Canada, 
        {\tt\small sean.lawlor@mail.mcgill.ca}}%
\thanks{M.G. Rabbat is an Assistant Professor at the Department of Electrical and Computer Engineering, McGill University, Montreal, Quebec H3A 2A7, Canada, 
        {\tt\small michael.rabbat@mcgill.ca}}%
}

\begin{document}

\maketitle

\begin{abstract}
We study the scalability of consensus-based distributed optimization algorithms by considering two questions: How many processors should we use for a given problem, and how often should they communicate when communication is not free? Central to our analysis is a problem-specific value $r$ which quantifies the communication/computation tradeoff. We show that organizing the communication among nodes as a $k$-regular expander graph~\cite{kRegExpanders} yields speedups, while when all pairs of nodes communicate (as in a complete graph), there is an optimal number of processors that depends on $r$. Surprisingly, a speedup can be obtained, in terms of the time to reach a fixed level of accuracy, by communicating less and less frequently as the computation progresses. Experiments on a real cluster solving metric learning and non-smooth convex minimization tasks demonstrate strong agreement between theory and practice. 
\end{abstract}




\section{Introduction}
How many processors should we use and how often should they communicate for large-scale distributed optimization? We address these questions by studying the performance and limitations of a class of distributed algorithms that solve the general optimization problem 
\begin{align} \label{eq:general_optimization_problem}
\underset{x \in \mathcal{X}}{\operatorname{minimize}}\ F(x) = \frac{1}{m} \sum_{j=1}^m l_j(x)
\end{align}
where each function $l_j(x)$ is convex over a convex set $\mathcal{X} \subseteq \mathds{R}^d$. This formulation applies widely in machine learning scenarios, where $l_j(x)$ measures the loss of model $x$ with respect to data point $j$, and $F(x)$ is the cumulative loss over all $m$ data points.

Although efficient serial algorithms exist \cite{nesterovDualOpt}, the increasing size of available data and problem dimensionality are pushing computers to their limits and the need for parallelization arises \cite{langfordBook}. Among many proposed distributed approaches for solving \eqref{eq:general_optimization_problem}, we focus on consensus-based distributed optimization \cite{dualAveraging,nedicDistributedOptimization,JohanssonIncrementalSubgrad,distrStochSubgrOpt} where each component function in \eqref{eq:general_optimization_problem} is assigned to a different node in a network (i.e., the data is partitioned among the nodes), and the nodes interleave local gradient-based optimization updates with communication using a consensus protocol to collectively converge to a minimizer of $F(x)$. 

Consensus-based algorithms are attractive because they make distributed optimization possible without requiring centralized coordination or significant network infrastructure (as opposed to, e.g., hierarchical schemes~\cite{delayedDistrOpt}). In addition, they combine simplicity of implementation with robustness to node failures and are resilient to communication delays \cite{TsianosACC2012}. These qualities are important in clusters, which are typically shared among many users, and algorithms need to be immune to slow nodes that use part of their computation and communication resources for unrelated tasks. The  main drawback of consensus-based optimization algorithms comes from the potentially high communication cost associated with distributed consensus.  At the same time, existing convergence bounds in terms of iterations (e.g., \eqref{eq:dda_error} below) suggest that increasing the number of processors slows down convergence, which contradicts the intuition that more computing resources are better.

This paper focuses on understanding the limitations and potential for scalability of consensus-based optimization. We build on the distributed dual averaging framework~\cite{dualAveraging}. The key to our analysis is to attach to each iteration a cost that involves two competing terms: a computation cost per iteration which decreases as we add more processors, and a communication cost which depends on the network. Our cost expression quantifies the communication/computation tradeoff by a parameter $r$ that is easy to estimate for a given problem and platform. The role of $r$ is essential; for example, when nodes communicate at every iteration, we show that in complete graph topologies, there exists an optimal number of processors $n_{opt} = \frac{1}{\sqrt{r}}$, while for $k$-regular expander graphs \cite{kRegExpanders}, increasing the network size yields a diminishing speedup. Similar results are obtained when nodes communicate every $h>1$ iterations and even when $h$ increases with time. We validate our analysis with experiments on a cluster. Our results show a remarkable agreement between theory and practice. 

In Section \ref{sec:distributed_optimization} we formalize the distributed optimization problem and summarize the distributed dual averaging algorithm. Section \ref{sec:comm_comp_tradeoff} introduces the communication/computation tradeoff and contains the basic analysis where nodes communicate at every iteration. The general case of sparsifying communication is treated in Section \ref{sec:sparse_communication}. Section \ref{sec:experiments} tests our theorical results on a real cluster implementation and Section \ref{sec:conclusions} discusses some future extensions.

\section{Distributed Convex Optimization}
\label{sec:distributed_optimization}

Assume we have at our disposal a cluster with $n$ processors to solve~\eqref{eq:general_optimization_problem}, and suppose without loss of generality that $m$ is divisible by $n$. In the absence of any other information, we partition the data evenly among the processors and our objective becomes to solve the optimization problem,
\begin{align} \label{eq:distributed_optimization_problem}
\underset{x \in \mathcal{X}}{\operatorname{minimize}}\ F(x) = \frac{1}{m} \sum_{j=1}^m l_j(x) = \frac{1}{n} \sum_{i=1}^n \left( \frac{n}{m} \sum_{j=1}^{\frac{m}{n}} l_{j|i}(x) \right) = \frac{1}{n} \sum_{i=1}^n f_i(x)
\end{align}
where we use the notation $\l_{j|i}$ to denote loss associated with the $j$th local data point at processor $i$ (i.e., $j|i = (i-1)\frac{m}{n} + j$). The local objective functions $f_i(x)$ at each node are assumed to be $L$-Lipschitz and convex. The recent distributed optimization literature contains multiple consensus-based algorithms with similar rates of convergence for solving this type of problem. We adopt the \emph{distributed dual averaging} (DDA) framework~\cite{dualAveraging} because its analysis admits a clear separation between the standard (centralized) optimization error and the error due to distributing computation over a network, facilitating our investigation of the communication/computation tradeoff.

\subsection{Distributed Dual Averaging (DDA)}
\label{sec:dda}

In DDA, nodes iteratively communicate and update optimization variables to solve~\eqref{eq:distributed_optimization_problem}. Nodes only communicate if they are neighbors in a communication graph $G = (V,E)$, with the $|V|=n$ vertices being the processors. The communication graph is user-defined (application layer) and does not necessarily correspond to the physical interconnections between processors.
DDA requires three additional quantities: a $1$-strongly convex proximal function $\psi: \mathds{R}^{d} \rightarrow \mathds{R}$ satisfying $\psi(x) \geq 0$ and $\psi(0) = 0$ (e.g., $\psi(x) = \frac{1}{2} x^T x$); a positive step size sequence $a(t) = O(\frac{1}{\sqrt{t}})$; and a $n \times n$ doubly stochastic consensus matrix $P$ with entries  $p_{ij} > 0$ only if either $i=j$ or $(j,i) \in E$ and $p_{ij} = 0$ otherwise. The algorithm repeats for each node $i$ in discrete steps $t$,  the following updates:
\begin{align} 
z_{i}(t) = &\sum_{j=1}^{n} p_{ij} z_{j}(t-1) + g_{i}(t-1) \label{eq:dda_z} \\
x_{i}(t) = & \underset{x \in \mathcal{X}}{\operatorname{argmin}} \left\lbrace  \langle z_i(t), x \rangle + \frac{1}{a(t)} \psi(x) \right\rbrace \label{eq:dda_x} \\
\hat{x}_{i}(t) = &  \frac{1}{t} \big( (t-1) \cdot \hat{x}_i(t-1)  + x_i(t) \big) \label{eq:dda_xhat}
\end{align}
where $g_i(t-1) \in \partial f_i(x_i(t-1))$ is a subgradient of $f_i(x)$ evaluated at $x_i(t-1)$.  In \eqref{eq:dda_z}, the variable $z_{i}(t) \in \mathds{R}^d$ maintains an accumulated subgradient up to time $t$ and represents node $i$'s belief of the direction of the optimum. To update $z_{i}(t)$ in~\eqref{eq:dda_z}, each node must communicate to exchange the variables $z_j(t)$ with its neighbors in $G$. If $\psi(x^*) \leq R^2$, for the local running averages $\hat{x}_{i}(t)$ defined in~\eqref{eq:dda_xhat}, the error from a minimizer $x^*$ of $F(x)$  after $T$ iterations is bounded  by (Theorem $1$, \cite{dualAveraging})
\begin{align} \label{eq:dda_basic_bound}
\text{Err}_i(T) = F(\hat{x}_i(T)) - F(x^*) \leq & \frac{R^2}{T a(T)} + \frac{L^2}{2T} \sum_{t=1}^T a(t-1) \notag\\
&+ \frac{L}{T} \sum_{t=1}^T a(t) \left( \frac{2}{n} \sum_{j=1}^n \norm{\bar{z}(t) - z_j(t)}_* + \norm{\bar{z}(t) - z_i(t)}_* \right)
\end{align}
where $L$ is the Lipschitz constant, $\norm{\cdot}_*$ indicates the dual norm, $\bar{z}(t) = \frac{1}{n} \sum_{i=1}^n z_i(t)$, and $\norm{\bar{z}(t) - z_i(t)}_*$ quantifies the network error as a disagreement between the direction to the optimum at node $i$ and the consensus direction $\bar{z}(t)$ at time $t$.  Furthermore, from Theorem $2$ in \cite{dualAveraging}, with $a(t) = \frac{A}{\sqrt{t}}$, after optimizing for $A$ we have a bound on the error,
\begin{align} \label{eq:dda_error}
\text{Err}_i(T)  \leq  C_1 \frac{\log{(T\sqrt{n})}}{\sqrt{T}},\quad\quad C_1 = 2 L R \sqrt{19 + \frac{12}{1- \sqrt{\lambda_2}}},
\end{align}
where $\lambda_2$ is the second largest eigenvalue of $P$. The dependence on the communication topology is reflected through $\lambda_2$, since the sparsity structure of $P$ is determined by $G$. According to~\eqref{eq:dda_error}, increasing $n$ slows down the rate of convergence even if $\lambda_2$  does not depend on $n$.

\section{Communication/Computation Tradeoff}
\label{sec:comm_comp_tradeoff}

In consensus-based distributed optimization algorithms such as DDA, the communication graph $G$ and the cost of transmitting a message have an important influence on convergence speed, especially when communicating one message requires a non-trivial amount of time (e.g., if the dimension $d$ of the problem is very high). 

We are interested in the shortest time to obtain an $\epsilon$-accurate solution (i.e., $\text{Err}_i(T) \leq \epsilon$). From~\eqref{eq:dda_error}, convergence is faster for topologies with good expansion properties; i.e., when the spectral gap $1 - \sqrt{\lambda_2}$ does not shrink too quickly as $n$ grows. In addition, it is preferable to have a balanced network, where each node has the same number of neighbors so that all nodes spend roughly the same amount of time communicating per iteration. Below we focus on two particular cases and take $G$ to be either a complete graph (i.e., all pairs of nodes communicate) or a $k$-regular expander~\cite{kRegExpanders}. 

By using more processors, the total amount of communication inevitably increases. At the same time, more data can be processed in parallel in the same amount of time. We focus on the scenario where the size $m$ of the dataset is fixed but possibly very large. To understand whether there is room for speedup, we move away from measuring iterations and employ a time model that explicitly accounts for communication cost. This will allow us to study the communication/computation tradeoff and draw conclusions based on the total amount of time to reach an $\epsilon$ accuracy solution. 

\subsection{Time model}

At each iteration, in step~\eqref{eq:dda_z}, processor $i$ computes a local subgradient on its subset of the data:
\begin{align}
g_{i}(x) = \frac{\partial{f_{i}(x)}}{\partial{x}} = \frac{n}{m} \sum_{j=1}^{\frac{m}{n}}  \frac{\partial{l_{j|i}(x)}}{\partial{x}}.
\end{align}
The cost of this computation increases linearly with the subset size. Let us normalize time so that one processor compute a subgradient on the full dataset of size $m$ in $1$ time unit. Then, using $n$ cpus, each local gradient will take $\frac{1}{n}$ time units to compute. We ignore the time required to compute the projection in step~\eqref{eq:dda_x}; often this can be done very efficiently and requires negligible time when $m$ is large compared to $n$ and $d$.

We account for the cost of communication as follows. In the consensus update~\eqref{eq:dda_z}, each pair of neighbors in $G$ transmits and receives one variable $z_j(t-1)$. Since the message size depends only on the problem dimension $d$ and does not change with $m$ or $n$, we denote by $r$ the time required to transmit and receive one message, relative to the $1$ time unit required to compute the full gradient on all the data. If every node has $k$ neighbors, the cost of one iteration in a network of $n$ nodes is 
\begin{align}
\frac{1}{n} + k r \text{ \ time units / iteration}.
\end{align}
Using this time model, we study the convergence rate bound~\eqref{eq:dda_error} after attaching an appropriate time unit cost per iteration. To obtain a speedup by increasing the number of processors $n$ for a given problem, we must ensure that $\epsilon$-accuracy is achieved in fewer time units. 

\subsection{Simple Case: Communicate at every Iteration}

In the original DDA description \eqref{eq:dda_z}-\eqref{eq:dda_xhat}, nodes communicate at every iteration. According to our time model, $T$ iterations will cost $\tau =  T (\frac{1}{n} + k r ) $ time units. From \eqref{eq:dda_error}, the time $\tau(\epsilon)$ to reach error $\epsilon$ is found by substituting for $T$ and solving for $\tau(\epsilon)$. Ignoring the log factor in~\eqref{eq:dda_error}, we get
\begin{align} \label{eq:dda_timeunit_error}
C_1 \frac{ 1}{\sqrt{\frac{\tau(\epsilon)}{\frac{1}{n} + k r}}} = \epsilon \quad \Longrightarrow \quad \tau(\epsilon) = \frac{C_1^{2}}{\epsilon^{2}} \Big( \frac{1}{n} + k r\Big) \text{ time units}.
\end{align}
This simple manipulation reveals some important facts. If communication is free, then $r=0$. If in addition the network $G$ is a  $k$-regular expander, then $\lambda_2$  is fixed \cite{SpectralGraphTheoryBook}, $C_1$ is independent of $n$ and $\tau(\epsilon) = C_1^{2} / ( \epsilon^{2} n)$. Thus, in the ideal situation, we obtain a linear speedup by increasing the number of processors, as one would expect. In reality, of course, communication is not free. 

\textbf{Complete graph.} Suppose that $G$ is the complete graph, where $k = n-1$ and $\lambda_2 = 0$. In this scenario we cannot keep increasing the network size without eventually harming performance due to the excessive communication cost. For a problem with a communication/computation tradeoff $r$, the optimal number of processors  is calculated by minimizing $\tau(\epsilon)$ for $n$:
\begin{align} \label{eq:optimal_numCPU}
\frac{\partial{\tau(\epsilon)}}{\partial{n}} = 0 \quad \Longrightarrow \quad n_{opt} =  \frac{1}{ \sqrt{r}}.
\end{align}
Again, in accordance with intuition, if the communication cost is too high (i.e., $r \geq 1$) and it takes more time to transmit and receive a gradient than it takes to compute it, using a complete graph cannot speedup the optimization. We reiterate that $r$ is a quantity that can be easily measured for a given hardware and a given optimization problem. As we report in Section \ref{sec:experiments}, the optimal value predicted by our theory agrees very well with experimental performance on a real cluster.

\textbf{Expander.} For the case where $G$ is a $k$-regular expander, the communication cost per node remains constant as $n$ increases. From \eqref{eq:dda_timeunit_error} and the expression for $C_1$ in~\eqref{eq:dda_error}, we see that $n$ can be increased without losing performance, although the benefit diminishes (relative to $kr$) as $n$ grows.

\section{General Case: Sparse Communication}
\label{sec:sparse_communication}

The previous section analyzes the case where processors communicate at every iteration. 
Next we investigate the more general situation where we adjust the frequency of communication.

\subsection{Bounded Intercommunication Intervals}

Suppose that a consensus step takes place once every $h+1$ iterations. That is, the algorithm repeats $h \geq 1$ cheap iterations (no communication) of cost $\frac{1}{n}$ time units followed by an expensive iteration (with communication) with cost $\frac{1}{n} + k r$. This strategy clearly reduces the overall average cost per iteration. The caveat is that the network error $\norm{\bar{z}(t) - z_i(t)}_*$ is higher because of having executed fewer consensus steps.

In a cheap iteration we replace the update \eqref{eq:dda_z} by $z_i(t) = z_i(t-1) + g_i(t-1)$. After some straight-forward algebra we can show that [for \eqref{eq:proof_zi_boundedcomm}, \eqref{eq:newNetErrorBound} please consult the supplementary material]:
\begin{align} \label{eq:proof_zi_boundedcomm}
z_{i}(t) & = \sum_{w=0}^{ H_{t} - 1} \sum_{k=0}^{h-1} \sum_{j=1}^{n} \left[ P^{ H_{t}- w} \right]_{ij}   g_{j}(w h + k) + \sum_{k=0}^{Q_t - 1} g_{i}(t - Q_t + k).
\end{align}
where $H_t = \lfloor \frac{t-1}{h}\rfloor$ counts the number of communication steps in $t$ iterations, and $Q_t = \text{mod}(t,h)$ if $\text{mod}(t,h) > 0$ and $Q_t = h$ otherwise. Using the fact that $P \vc{1} = \vc{1}$, we obtain
\begin{align}
\bar{z}(t) - z_i(t) = \frac{1}{n} \sum_{s=1}^n z_s(t) - z_i(t) =&  \sum_{w=0}^{ H_{t} - 1} \sum_{j=1}^{n} \Big( \frac{1}{n} - \left[ P^{ H_{t}- w} \right]_{ij} \Big)  \sum_{k=0}^{h-1}  g_{j}(w h + k) \\
& + \frac{1}{n } \sum_{s=1}^{n} \sum_{k=0}^{Q_t - 1} \big(g_{s}(t - Q_t + k) - g_{i}(t - Q_t + k) \big).
\end{align}
Taking norms, recalling that the $f_i$ are convex and Lipschitz, and since $Q_t \leq h$, we arrive at
\begin{align} \label{eq:network_error_norm}
\norm{\bar{z}(t) - z_i(t)}_* \leq  \sum_{w=0}^{ H_{t} - 1} \norm{  \frac{1}{n} \vc{1}^T - \left[ P^{ H_{t}- w} \right]_{i,:} }_1  h L +  2 h L 
\end{align}
Using a technique similar to that in~\cite{dualAveraging} to bound the $\ell_1$ distance of row $i$ of $P^{H_t - w}$ to its stationary distribution as $t$ grows, we can show that
\begin{align} \label{eq:newNetErrorBound}
\norm{\bar{z}(t) - z_{i}(t)}_{*} \leq   2 h L \frac{\log(T \sqrt{n})}{1 - \sqrt{\lambda_{2}}} + 3 h L
\end{align}
for all $t \leq T$.  Comparing~\eqref{eq:newNetErrorBound} to equation (29) in \cite{dualAveraging}, the network error within $t$ iterations is no more than $h$ times larger when a consensus step is only performed once every $h+1$ iterations. Finally, we substitute the network error in \eqref{eq:dda_basic_bound}. For $a(t) = \frac{A}{\sqrt{t}}$, we have $\sum_{t=1}^T a(t) \leq 2 A \sqrt{T}$, and
\begin{align}
\text{Err}_i(T) \leq \left( \frac{R^2}{A} +  A L^2  \left(1  + \frac{12 h    }{1 - \sqrt{\lambda_2}} + 18h \right)\right)  \frac{\log{(T\sqrt{n})}}{\sqrt{T}} = C_h  \frac{\log{(T\sqrt{n})}}{\sqrt{T}} .
\end{align}
We minimize  the leading term $C_h$ over $A$ to obtain 
\begin{align}
A = \frac{R}{L} \left( \sqrt{1 + 18h + \frac{12 h }{1 - \sqrt{\lambda_{2}}}} \right)^{-1} \text{\ and\ \ \ } C_h = 2RL \sqrt{1 + 18h + \frac{12 h }{1 - \sqrt{\lambda_{2}}}}.
\end{align}

Of the $T$ iterations, only $H_{T} = \lfloor \frac{T-1}{h} \rfloor$  involve communication. So, $T$ iterations will take
\begin{align}
\tau = (T - H_{T}) \frac{1}{n} + H_{T} \left( \frac{1}{n} + k r \right) = \frac{T}{n} + H_{T} k r \text{\ \ time\ units}.
\end{align}
To achieve $\epsilon$-accuracy, ignoring again the logarithmic factor, we need $T =  \frac{C_h^{2}}{\epsilon^{2}}$ iterations, or 
\begin{align}
\tau(\epsilon) = \left( \frac{T}{n} + \left\lfloor \frac{T-1}{h} \right\rfloor  kr \right) \leq \frac{C_h^{2}}{\epsilon^{2}} \left(\frac{1}{n} + \frac{kr}{h} \right) \text{ time units}.
\end{align}
From the last expression, for a fixed number of processors $n$, there exists an optimal value for $h$ that depends on the network size and communication graph $G$:
\begin{align} \label{eq:hopt}
h_{opt} = \sqrt{\frac{n k r}{18 + \frac{12}{1 - \sqrt{\lambda_2}}}}.
\end{align}
If the network is a complete graph, using $h_{opt}$ yields $\tau(\epsilon) = O(n)$; i.e., using more processors hurts performance when not communicating every iteration. On the other hand, if the network is a $k$-regular expander then $\tau(\epsilon) = \frac{c_1}{\sqrt{n}} + c_2$ for constants $c_1,c_2$, and we obtain a diminishing speedup.

\subsection{Increasingly Sparse Communication}

Next, we consider progressively increasing the intercommunication intervals. This captures the intuition that as the optimization moves closer to the solution, progress slows down and a processor should have ``something significantly new to say" before it communicates. Let $h_j-1$ denote the number of cheap iterations performed between the $(j-1)$st and $j$th expensive iteration; i.e., the first communication is at iteration $h_1$, the second at iteration $h_1 + h_2$, and so on. We consider schemes where $h_j = j^p$ for $p \ge 0$. The number of iterations that nodes communicate out of the first $T$ total iterations is given by $H_T = \max \{H \colon \sum_{j=1}^H h_j \le T\}$.
We have
\begin{align}
\int_{y=1}^{H_T}  y^p dy  \leq \sum_{j=1}^{H_{T}} j^{p}  & \leq 1 + \int_{y=1}^{H_T}  y^p dy \quad \Longrightarrow \quad \frac{H_T^{p+1} - 1}{p+1}  \leq T \leq  \frac{H_T^{p+1} + p}{p+1}, 
\end{align}
which means that $H_T = \Theta(T^{\frac{1}{p+1}})$ as $T \rightarrow \infty$. Similar to \eqref{eq:network_error_norm}, the network error is bounded as
\begin{align}
\norm{\bar{z}(t) - z_{i}(t)}_{*} \leq & \sum_{w=0}^{H_t - 1} \norm{  \frac{1}{n} \vc{1}^T - \left[ P^{ H_{t}- w} \right]_{i,:} }_1 \sum_{k=0}^{h_w-1} L + 2 h_t L = L \sum_{w=0}^{H_t - 1} \norm{\cdot}_1 h_w+ 2 h_t L.
\end{align}
We split the sum into two terms based on whether or not the powers of $P$ have converged. Using the split point $\hat{t} = \frac{\log(T\sqrt{n})}{1 - \sqrt{\lambda_{2}}}$, the $\ell_{1}$ term is bounded by $2$ when $w$ is large and by $\frac{1}{T}$ when $w$ is small: 
\begin{align}
\norm{\bar{z}(t) - z_{i}(t)}_{*} \leq & L \sum_{w=0}^{H_{t}-1 - \hat{t} } \norm{\cdot}_{1} h_{w} + L \sum_{w=H_{t} -\hat{t}}^{H_{t}-1 } \norm{\cdot}_{1} h_{w} + 2 h_{t} L \\
\leq & \frac{L}{T} \sum_{w=0}^{H_{t}-1 - \hat{t} } w^{p} + 2 L \sum_{w=H_{t} -\hat{t}}^{H_{t}-1 } w^{p} + 2 t^{p} L \\
\leq &  \frac{L}{T}  \frac{(H_t - \hat{t} - 1)^{\frac{1}{p+1}} + p}{p+1}  + 2L \hat{t} (H_t-1)^p + 2 t^p L \\
\leq &  \frac{L}{p+1} +  \frac{Lp}{T(p+1)}   + 2L \hat{t} H_t^p + 2 t^p L
\end{align}
since $T > H_t - \hat{t} - 1$. Substituting this bound into \eqref{eq:dda_basic_bound} and taking the step size sequence to be $a(t) = \frac{A}{t^q}$ with $A$ and $q$ to be determined, we get
\begin{align}
\text{Err}_i(T) \leq &\frac{R^2}{A T^{1-q}} + \frac{L^2 A}{2(1-q)T^q} + \frac{3L^2A}{(p+1)(1-q)T^q} + \frac{3L^2 p A}{(p+1)(1-q) T^{1+q}} \notag \\
& + \frac{6L^2 \hat{t} A}{T} \sum_{t=1}^{T} \frac{H_t^p}{t^q} + \frac{6L^2A}{T} \sum_{t=1}^T t^{p - q}.
\end{align}
The first four summands converge to zero when $0 < q < 1$. Since $H_t = \Theta(t^{\frac{1}{p+1}})$,
\begin{align}
\frac{1}{T} \sum_{t=1}^T \frac{H_t^p}{t^q} \leq  \frac{1}{T} \sum_{t=1}^T   \frac{ O(t^\frac{1}{p+1})^p}{t^q} \leq O \left( \frac{T^{\frac{p}{p+1} - q + 1}}{T} \right) =  O \left( T^{\frac{p}{p+1}-q} \right)
\end{align}
which converges to zero if $\frac{p}{p+1} < q$. To bound the last term, note that $\frac{1}{T} \sum_{t=1}^T t^{p-q} \leq \frac{T^{p-q}}{p-q+1}$,
so the term goes to zero as $T\rightarrow \infty$ if $p < q$. In conclusion, $\text{Err}_i(T)$ converges no slower than $O(\frac{\log{(T\sqrt{n})}}{T^{q-p}})$ since $\frac{1}{T^{q - \frac{p}{p+1}}} < \frac{1}{T^{q-p}}$. If we choose $q = \frac{1}{2}$ to balance the first three summands, for small $p > 0$, the rate of convergence is arbitrarily close to $O(\frac{\log{(T\sqrt{n})}}{\sqrt{T}})$, while nodes communicate increasingly infrequently as $T \rightarrow \infty$.

Out of $T$ total iterations, DDA executes $H_T = \Theta(T^{\frac{p}{p+1}})$ expensive iterations involving communication and $T- H_T$ cheap iterations without communication, so
\begin{align}
\tau(\epsilon) = O \left (\frac{T}{n} + T^{\frac{p}{p+1}} k r \right) = O \left( T \left(\frac{1}{n} + \frac{kr}{T^{\frac{1}{p+1}}} \right)\right).
\end{align}
In this case, the communication cost $kr$ becomes a less and less significant proportion of $\tau(\epsilon)$ as $T$ increases. So for any $0 < p < \frac{1}{2}$, if $k$ is fixed, we approach a linear speedup behaviour $\Theta(\frac{T}{n})$. To get $\text{Err}_i(T) \le \epsilon$, ignoring the logarithmic factor, we need
\begin{align}
T = \left(  \frac{C_p}{\epsilon} \right)^{\frac{2}{1 - 2p}} \text{ iterations, with } C_p = 2 L R \sqrt{7 + \frac{12 p + 12}{(3 p + 1)(1 - \sqrt{\lambda_2})}  + \frac{12}{2p+1} }.
\end{align}
From this last equation we see that for $0 < p < \frac{1}{2}$ we have $C_p < C_1$, so using increasingly sparse communication should, in fact, be faster than communicating at every iteration.

\section{Experimental Evaluation}
\label{sec:experiments}

To verify our theoretical findings, we implement DDA on a cluster of $14$ nodes with 3.2~GHz Pentium $4$HT processors and $1$ GB of memory each, connected via ethernet that allows for roughly $11$~MB/sec throughput per node. Our implementation is in C++ using the send and receive functions of OpenMPI v1.4.4 for communication. The Armadillo v2.3.91 library, linked to LAPACK and BLAS, is used for efficient numerical computations.

\subsection{Application to Metric Learning}
Metric learning \cite{metricLearnCluster,metricLearnNNclassification,MetricLearnConvexOpt} is a computationally intensive problem where the goal is to find a distance metric $D(u,v)$ such that points that are related have a very small distance under  $D$ while for unrelated points $D$ is large. Following the formulation in \cite{metricLearnOnlineBatch}, we have a data set $\{u_j, v_j, s_j \}_{j=1}^m$ with $u_j, v_j \in \mathds{R}^d$ and $s_j = \{-1, 1\}$ signifying whether or not $u_j$ is similar to $v_j$ (e.g., similar if they are from the same class). Our goal is to find a symmetric positive semi-definite matrix $A \succeq 0$ to define a pseudo-metric of the form $ D_A(u,v) = \sqrt{(u-v)^T A (u-v)}$.
To that end, we use a hinge-type loss function $ l_j(A,b) = \max\lbrace 0, s_j \left( D_A(u_j, v_j)^2  - b \right) + 1\rbrace$
where $b \geq 1$ is a threshold that determines whether two points are dissimilar according to $D_A(\cdot,\cdot)$. In the batch setting, we formulate the  convex optimization problem
\begin{align} \label{eq:metric_learning_cost}
\underset{A ,b} {\text{minimize}}\quad F(A,b) = \sum_{j=1}^m l_j(A,b)  \quad \text{subject to}  \quad A \succeq 0, b \geq 1.
\end{align}
The subgradient of $l_j$ at $(A,b)$ is zero if $s_j(D_A(u_j, v_j)^2 - b) \leq -1 $. Otherwise
\begin{align} \label{eq:metricLearningGrad}
\frac{\partial l_j(A,b)}{\partial A} = s_j(u_j - v_j)^T (u_j - v_j), \quad \text{and}\quad \frac{\partial l_j(A,b)}{\partial b} =  - s_j.
\end{align}
Since DDA uses vectors $x_i(t)$ and $z_i(t)$, we represent each pair $(A_i(t),b_i(t))$ as a $d^2+1$ dimensional vector. The communication cost is thus quadratic in the dimension. In step \eqref{eq:dda_z} of DDA, we use the proximal function $\psi(x) = \frac{1}{2} x^T x$, in which case~\eqref{eq:dda_x} simplifies to taking $x_i(t) = -a(t-1) z_i(t)$, followed by projecting $x_i(t)$ to the constraint set by setting $b_i(t) \leftarrow \max\{1, b_i(t)\}$ and projecting $A_i(t)$ to the set of positive semi-definite matrices by first taking its eigenvalue decomposition and reconstructing $A_i(t)$ after forcing any negative eigenvalues to zero.

\begin{figure}[t]
\begin{center}
\includegraphics[width=2.6in,height=1.5in]{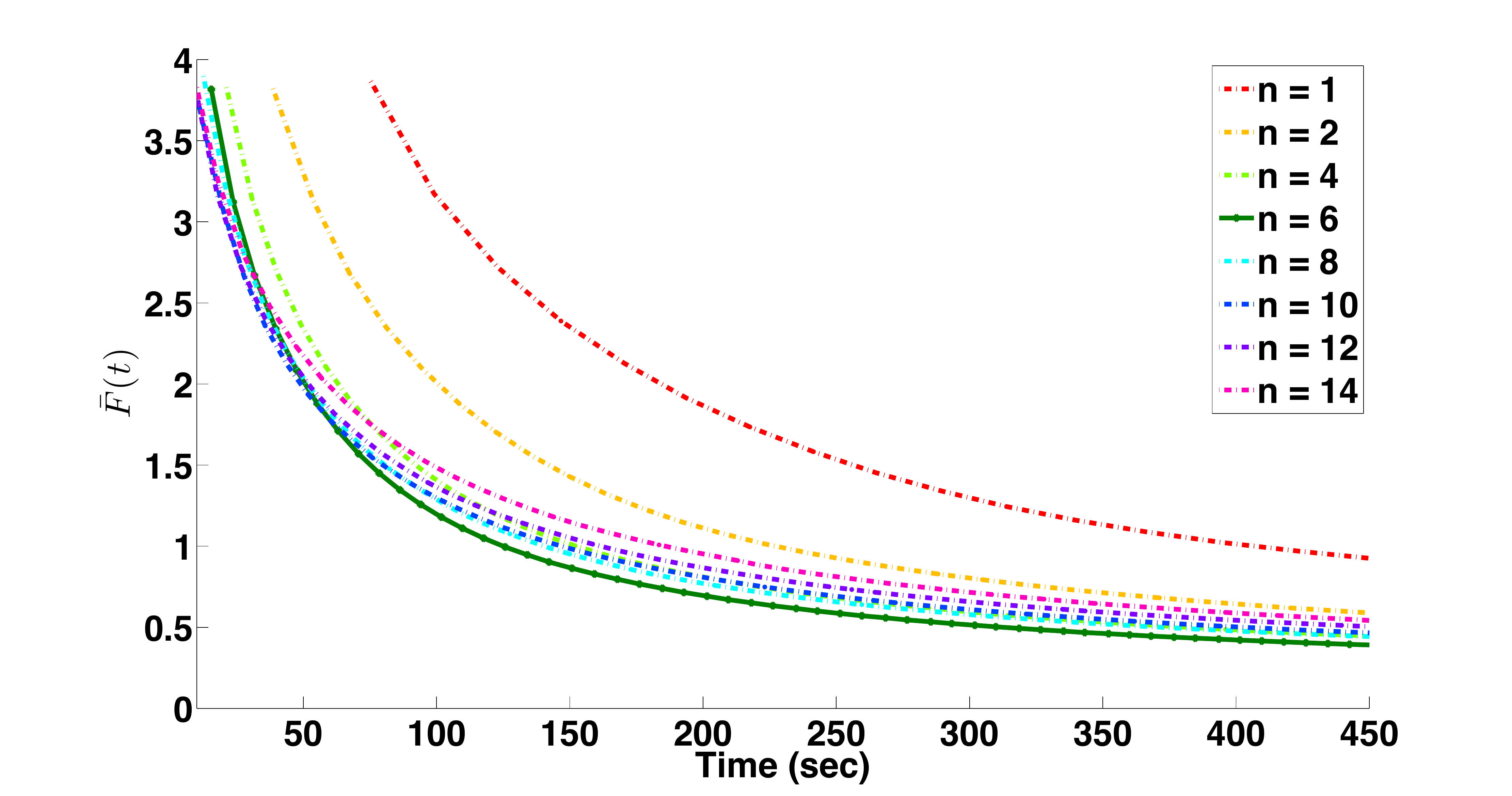} 
\includegraphics[width=2.6in,height=1.5in]{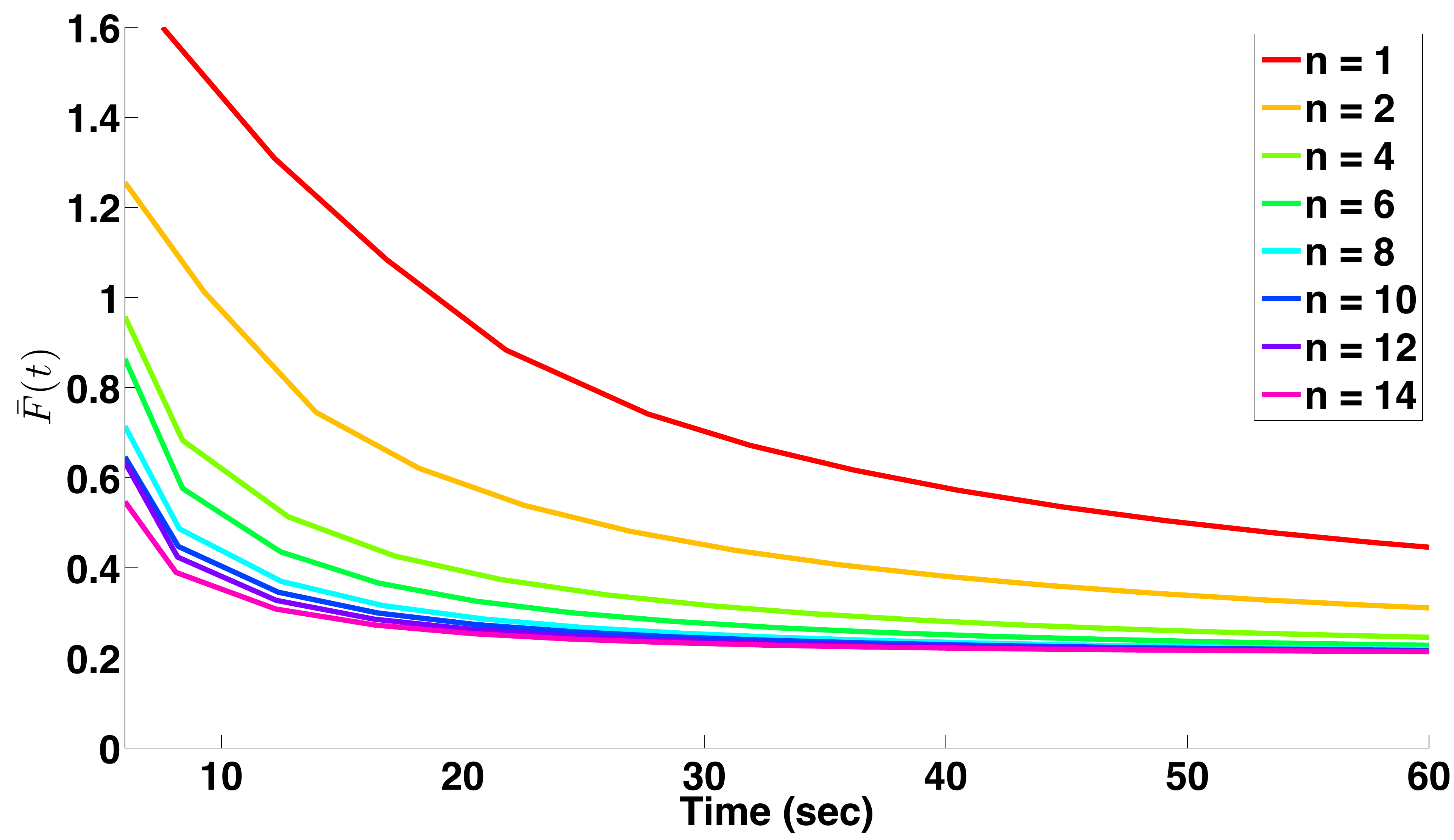} 
\end{center}
\caption{\label{fig:zeroCommCost} (Left) In a subset of the Full MNIST data for our specific hardware, $n _{opt} = \frac{1}{\sqrt{r}}= 5.8$. The fastest convergence is achieved on a complete graph of $6$ nodes. (Right) In the reduced MNIST data using PCA, the communication cost drops and a speedup is achieved by scaling up to $14$ processors.}
\end{figure}

We use the MNIST digits dataset which consists of $28 \times 28$ pixel images of handwritten digits $0$ through $9$. Representing images as vectors, we have $d = 28^2 = 784$ and a problem with $d^2+1 = 614657$ dimensions trying to learn a $784 \times 784$ matrix $A$.  With double precision arithmetic, each DDA message has a size approximately $4.7$ MB.  We construct a dataset by randomly selecting $5000$ pairs from the full MNIST data. One node needs $29$ seconds to compute a gradient on this dataset, and sending and receiving $4.7$ MB takes $0.85$ seconds. The communication/computation tradeoff value is estimated as $r = \frac{0.85}{29} \approx 0.0293$. According to \eqref{eq:optimal_numCPU}, when $G$ is a complete graph, we expect to have optimal performance when using $n_{opt} = \frac{1}{\sqrt{r}} = 5.8$ nodes. Figure \ref{fig:zeroCommCost}(left) shows the evolution of the average function value $\bar{F}(t) = \frac{1}{n}\sum_i F(\hat{x}_i(t))$ for  $1$ to $14$ processors connected as a complete graph, where $\hat{x}_i(t)$ is as defined in~\eqref{eq:dda_xhat}. There is a very good match between theory and practice since the fastest convergence is achieved with $n=6$ nodes. 

In the second experiment, to make $r$ closer to $0$, we apply PCA to the original data and keep the top $87$ principal components, containing $90\%$ of the energy. The dimension of the problem is reduced dramatically to $87 \cdot 87+1 = 7570$ and the message size to $59$ KB. Using $60000$ random pairs of MNIST data, the time to compute one gradient on the entire dataset with one node is $2.1$ seconds, while the time to transmit and receive $59$~KB is only $0.0104$ seconds. Again, for a complete graph, Figure \ref{fig:zeroCommCost}(right) illustrates the evolution of $\bar{F}(t)$ for $1$ to $14$ nodes. As we see, increasing $n$ speeds up the computation. The speedup we get is close to linear at first, but diminishes since communication is not entirely free. In this case $r = \frac{0.0104}{2.1} = 0.005$ and $n_{opt} = 14.15$.

\subsection{Nonsmooth Convex Minimization}
\begin{figure}[h]
\begin{center}
\includegraphics[width=2.6in,height=1.5in]{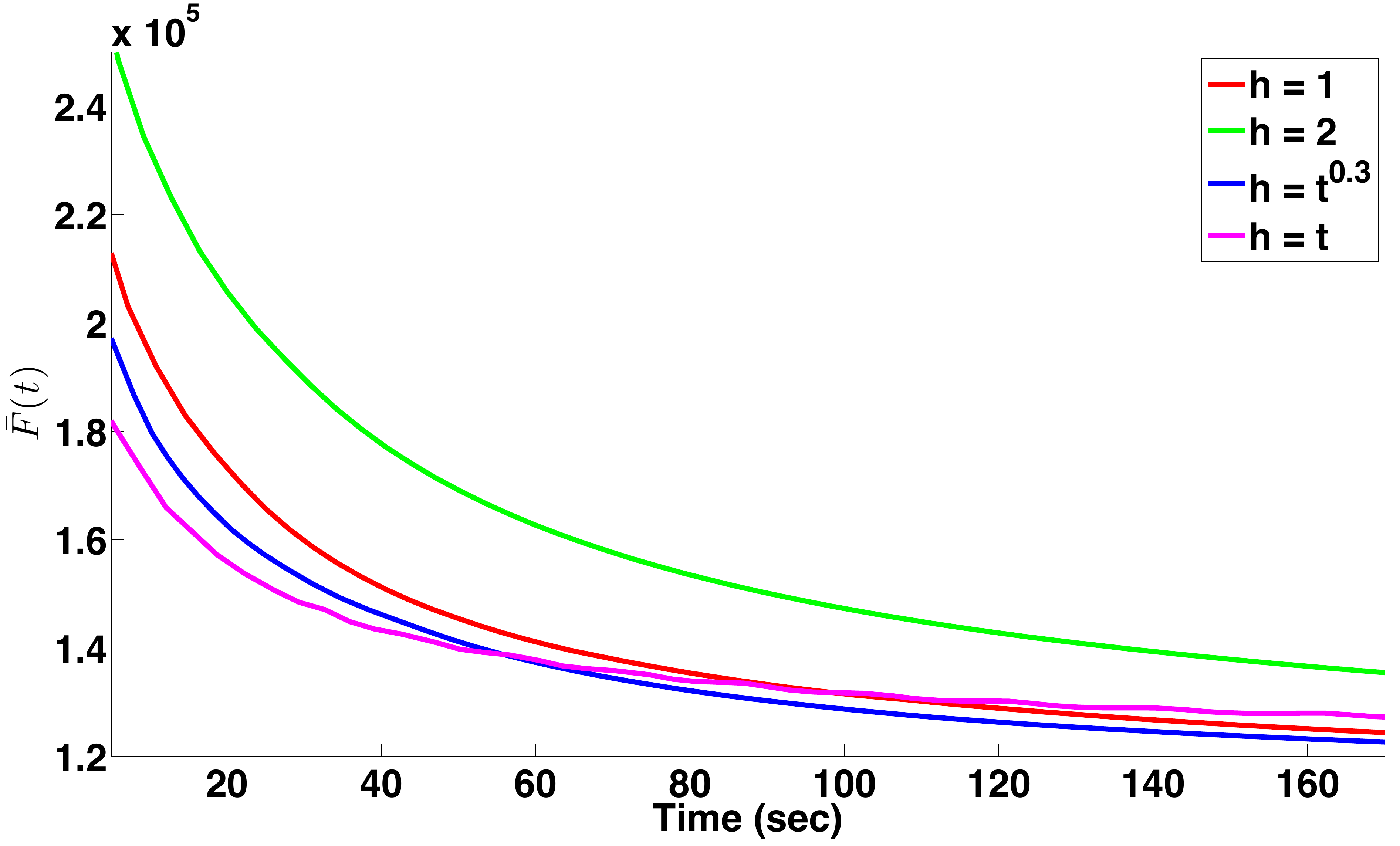}
\end{center}
\caption{\label{fig:SparsifyingComm} Sparsifying communication to minimize \eqref{eq:nonsmooth_convex} with $10$ nodes in a complete graph topology. When waiting $t^{0.3}$ iterations between consensus steps, convergence is faster than communicating at every iteration ($h=1$), even though the total number of consensus steps performed over the duration of the experiment is equal to communicating every $2$ iterations ($h=2$). When waiting a linear number of iterations between consensus steps ($h = t$) DDA does not converge to the right solution. Note: all methods are initialized from the same value; the x-axis starts at 5 sec.} 
\end{figure}

Next we create an artificial problem where the minima of the components $f_i(x)$ at each node are very different, so that communication is essential in order to obtain an accurate optimizer of $F(x)$. We define $f_i(x)$ as a sum of high dimensional quadratics, 
\begin{align} \label{eq:nonsmooth_convex}
f_i(x) =  \sum_{j=1}^{M} \operatorname{max} \left(l_{j|i}^1(x), l_{j|i}^2(x) \right), \quad l_{j|i}^\xi(x) = (x - c_{j|i}^\xi)^T (x - c_{j|i}^\xi), \quad \xi \in \{1,2\},
\end{align}
where $x \in \mathds{R}^{10,000}$, $M = 15,000$ and $c_{j|i}^{1}, c_{j|i}^{2}$ are the centers of the quadratics. Figure \ref{fig:SparsifyingComm} illustrates again the average function value $\bar{F}(t)$ for $10$ nodes in a complete graph topology. The baseline performance is when nodes communicate at every iteration ($h = 1$). For this problem $r = 0.00089$ and, from \eqref{eq:hopt},  $h_{opt} = 1$. Naturally communicating every $2$ iterations ($h=2$) slows down convergence. Over the duration of the experiment, with $h=2$,  each node communicates with its peers $55$ times. We selected $p=0.3$ for increasingly sparse communication, and got $H_T = 53$ communications per node. As we see,  even though nodes communicate as much as the $h=2$ case, convergence is even faster than communicating at every iteration. This verifies our intuition that communication is more important in the beginning. Finally, the case where $p=1$ is shown. This value is out of the permissible range, and as expected DDA does not converge to the right solution. 

\section{Conclusions and Future Work}
\label{sec:conclusions}


The analysis and experimental evaluation in this paper focus on distributed dual averaging and reveal the capability of distributed dual averaging to scale with the network size. We expect that similar results hold for other consensus-based algorithms such as \cite{nedicDistributedOptimization} as well as various distributed averaging-type algorithms (e.g., \cite{parallelSGD,distrPerceptron,distrMaxEntropyModels}). In the future we will extend the analysis to the case of stochastic optimization, where $h_t = t^p$ could correspond to using increasingly larger mini-batches. 

\newpage
\small{
\bibliographystyle{IEEEtran} 
\bibliography{../PhDThesis/References}
}

\newpage
\section{Appendix}

\subsection{Proof of equation \eqref{eq:proof_zi_boundedcomm}}
\label{app:proof_zi_boundedcomm}

Let us stack the local node variables in a vector $\vc{z} = [z_1 \cdots z_n]^T$ and $\vc{g} = [g_1 \cdots g_n]^T$. From \eqref{eq:dda_z} in matrix form we have after back-substituting in the recursion 
\begin{align}
\vc{z}(h+1) = P \vc{z}(h) + \vc{g}(h) = P \sum_{k=0}^{h-1} \vc{g}(k) + \vc{g}(h)
\end{align}
and after some algebra
\begin{align}
\vc{z}(s h + 1) = \sum_{w=1}^s \sum_{k=0}^{h-1}  P^w \vc{g}\big( (s-w) h + k\big) + \vc{g}(s h)
\end{align}
or in general
\begin{align}
\vc{z}(t) = &\sum_{w=1}^{H_t} \sum_{k=0}^{h-1}  P^w \vc{g}\big( (H_t - w) h + k\big) + \sum_{k=0}^{Q_t-1} \vc{g}(t - Q_t + k) \\
= & \sum_{w=1}^{H_t} \sum_{k=0}^{h-1}  P^{H_t - w + 1} \vc{g}\big( (w-1) h + k\big) + \sum_{k=0}^{Q_t-1} \vc{g}(t - Q_t + k)  \\
= & \sum_{w=0}^{H_t - 1} \sum_{k=0}^{h-1}  P^{H_t - w} \vc{g}\big(w h + k\big) + \sum_{k=0}^{Q_t-1} \vc{g}(t - Q_t + k) 
\end{align}
where $H_t = \lfloor \frac{t-1}{h}\rfloor$ counts the number of communication steps in $t$ iterations  and $Q_t = \text{mod}(t,h)$ if $\text{mod}(t,h) > 0$ and $Q_t = h$ otherwise.  From this last expression we take the $i$-th row to get the result.

\subsection{Proof of equation \eqref{eq:newNetErrorBound}}
\label{app:proof_zi_boundedcomm}

If the consensus matrix $P$ is doubly stochastic it is straightforward to show that $P^t \rightarrow \frac{1}{n} \vc{1} \vc{1}^T$ as $t \rightarrow \infty$. Moreover, from standard Perron-Frobenius is it easy to show (see e.g., \cite{StookDiaconis})
\begin{align}
\norm{\frac{1}{n} \vc{1}^T - \left[ P^t \right]_{i,:}}_1 = 2 \norm{\frac{1}{n} \vc{1}^T - \left[ P^t \right]_{i,:}}_{TV} \leq \sqrt{n} \left(\sqrt{\lambda_2}\right)^t
\end{align}
so in our case $\norm{\frac{1}{n} \vc{1}^T - \left[ P^{H_t - w} \right]_{i,:}}_1  \leq  \sqrt{n} \left(\sqrt{\lambda_2}\right)^{H_t - w}$. Next, demand that the right hand side bound is less than $\sqrt{n} \delta$ with $\delta$ to be determined:
\begin{align}
\sqrt{n} \left(\sqrt{\lambda_2}\right)^{H_t - w} \leq \sqrt{n} \delta \Rightarrow H_t - w \geq \frac{\log{(\delta^{-1})}}{\log{(\sqrt{\lambda_2}^{-1})}}.
\end{align}
So with the  choice $\delta^{-1} = \sqrt{n} T$,
\begin{align}
 \norm{\frac{1}{n} \vc{1}^T - \left[ P^{H_t - w} \right]_{i,:}}_1 \leq \sqrt{n} \frac{1}{\sqrt{n} T} = \frac{1}{T}
\end{align}
if $H_t - w \geq \frac{\log{(\delta^{-1})}}{\log{(\sqrt{\lambda_2}^{-1})}} = \hat{t}$. When $w$ is large and $H_t - w < \hat{t}$ we simply take $\norm{\frac{1}{n} \vc{1}^T - \left[ P^{H_t - w} \right]_{i,:}}_1 \leq 2$. The desired bound of \eqref{app:proof_zi_boundedcomm} is not obtained as follows
\begin{align}
\sum_{w=0}^{ H_{t} - 1} & \norm{  \frac{1}{n} \vc{1}^T - \left[ P^{ H_{t}- w} \right]_{i,:} }_1  h L +  2 h L  \notag \\
& = \left(\sum_{w=0}^{H_t - \hat{t} -1 } \norm{  \frac{1}{n} \vc{1}^T - \left[ P^{ H_{t}- w} \right]_{i,:} }_1 + \sum_{H_t - \hat{t}}^{H_t - 1} \norm{  \frac{1}{n} \vc{1}^T - \left[ P^{ H_{t}- w} \right]_{i,:} }_1\right) h L  +  2 h L \\
& \leq \left(\sum_{w=0}^{H_t - \hat{t} -1 } \frac{1}{T} + \sum_{H_t - \hat{t}}^{H_t - 1} 2 \right) h L  +  2 h L \\
& \leq \frac{H_t - \hat{t} }{T} h L  + 2 \hat{t} h L  +  2 h L.
\end{align}
Since $t < T$ we know that $H_t - \hat{t} < T$. Moreover, $\log{(\sqrt{\lambda_2})^{-1}} \geq 1 - \sqrt{\lambda_2}$. Using there two fact we arrive at the result.


\end{document}